# On Outage and Error Rate Analysis of the Ordered V-BLAST*

Sergey Loyka[1], Francois Gagnon[2]

*Abstract*- Outage and error rate performance of the ordered BLAST with more than 2 transmit antennas is evaluated for i.i.d. Rayleigh fading channels. A number of lower and upper bounds on the 1$^{st}$ step outage probability at any SNR are derived, which are further used to obtain accurate approximations to average block and total error rates. For *m* Tx antennas, the effect of the optimal ordering at the first step is an *m*-fold SNR gain. As *m* increases to infinity, the BLER decreases to zero, which is a manifestation of the space-time autocoding effect in the V-BLAST. While the sub-optimal ordering (based on the before-projection SNR) suffers a few dB SNR penalty compared to the optimal one, it has a lower computational complexity and a 3 dB SNR gain compared to the unordered V-BLAST and can be an attractive solution for low-complexity/low-energy systems. Uncoded D-BLAST exhibits the same outage and error rate performance as that of the V-BLAST. An SNR penalty of the linear receiver interfaces compared to the BLAST is also evaluated.

## I. INTRODUCTION

In recent years, the BLAST algorithm has gained significant popularity as a receiver interface for a spatial multiplexing system [1][2]. Its error rate performance has been initially studied using Monte-Carlo simulations and prototyped systems, and later on analytical techniques have been developed to attack the problem. Performance evaluation of the unordered V-BLAST can be done analytically by adapting the techniques developed for multiuser detection problems, resulting in exact closed-form expressions for average error rates and outage probabilities. This provides significant insight into algorithm's performance and its bottlenecks, and hence opening up opportunities for optimization [3]-[5][13][14]. However, the optimally ordered V-BLAST presents a serious challenge (due to the ordering procedure, which changes the channel statistics) for the analytical analysis. For the system with 2 transmit (Tx) antennas ( $m = 2$ ), exact closed-form expressions for outage and average error probabilities can be obtained for i.i.d. Rayleigh fading channel [6], but for $m \geq 3$ the problem complexity increases significantly and no exact closed-form expressions have been found so far. A diversity-order-based analysis of the V-BLAST was presented in [11][12]. In particular, it was demonstrated that the ordering procedure does not affect the diversity order. While this is a significant insight into the system performance, it does not give the complete picture since (i) it holds only asymptotically (as SNR $\to \infty$), and (ii) it does not say anything about the SNR-independent constant that may significantly affect the error rate performance, even at high SNR.

[1] The School of Information Technology and Engineering (SITE), University of Ottawa, 161 Louis Pasteur, Ottawa, Ontario, Canada, K1N 6N5 (sergey.loyka@ieee.org)
[2] Department of Electrical Engineering, Ecole de Technologie Superieure, 1100, Notre-Dame St. West, Montreal, Quebec, H3C 1K3, Canada (francois.gagnon@etsmtl.ca).

In this paper, we address this problem by deriving a number of bounds on the outage probability, which hold at any SNR, for the ordered zero-forcing (ZF) V-BLAST with more than 2 Tx antennas in the i.i.d. Rayleigh fading channel using a geometrically-based framework and a step-by-step analysis of the outage probability. This extends the corresponding results in [6] obtained for 2 Tx antennas. We demonstrate that one of the upper bounds is tight and, based on this, derive accurate closed-form approximations to the average block error rate (BLER)[3] and the total error rate (TBER)[4]. It is observed that the effect of the optimal ordering is an $m$-fold increase in 1$^{st}$ step after-processing SNR, or equivalently a $10\log m$ dB gain, which further extends to the $m$-fold SNR gain in terms of the average BLER[5]. Thus, while the optimal ordering does not provide any advantage in terms of the diversity gain, it does provide advantage in terms of the SNR gain, which increases with the number of Tx antennas. Based on these results, we conjecture that the ordered ZF V-BLAST possesses the autocoding effect originally discovered in [8] using information-theoretic arguments. A similar conjecture has been made in [5] for the unordered ZF V-BLAST in terms of the TBER (which does not hold in terms of the BLER in that case).

Based on the results above, the sub-optimal ordering (maximizing the before-projection SNR) is compared to the optimal ordering and also to the un-ordered V-BLAST. While the sub-optimal ordering suffers a few dB SNR penalty compared to the optimal ordering, it does provide a 3 dB SNR gain compared to the un-ordered V-BLAST and, hence, can be a low-complexity alternative to the optimally ordered V-BLAST.

The paper is organized as follows. Section II introduces a basic system and V-BLAST model and briefly reviews the relevant results in [3]-[6]. Section III generalizes the results in [6] to the case of more than 2 Tx antennas, gives a number of bounds on the outage probability, which allow to compare the optimally ordered V-BLAST to the unordered and sub-optimally ordered one, and quantifies the SNR gain of optimal and sub-optimal orderings. Sections IV gives high-SNR approximations for the average BLER and TBER respectively. Section V briefly discusses the error rate performance of the D-BLAST, compares the BLAST to the linear receiver interfaces and quantifies their SNR loss. Section VI concludes the paper.

---

[3] termed "frame error rate" in [9] and "joint error probability" in [3]. It is defined as a probability to have at least one error in the detected transmit symbol vector.

[4] termed "per-symbol error probability" in [3]. It is defined as the error rate at the output stream to which all the individual sub-streams are merged after the detection.

[5] While the diversity-order-based analysis in [9][11][12] does capture the diversity gain, it is not able to capture this SNR gain.

## II. CHANNEL MODEL AND V-BLAST ALGORITHM

The standard baseband system model is given by

$$\mathbf{r} = \mathbf{Hs} + \xi \qquad (1)$$

where $\mathbf{s}$ and $\mathbf{r}$ are the Tx and Rx vectors correspondingly, $\mathbf{H}$ is the $n \times m$ channel matrix, i.e. the matrix of the complex channel gains between each Tx and each Rx antenna, n is the number of Rx antennas, $m$ is the number of Tx antennas, $n \geq m$, and $\xi$ is the additive white Gaussian noise (AWGN), which is assumed to be $\mathcal{CN}(0, \sigma_0^2 \mathbf{I})$, i.e. independent and identically distributed (i.i.d.) in each branch. We assume i.i.d. Rayleigh fading channel, i.e. the entries of $\mathbf{H}$ are i.i.d. complex Gaussian with unit variance and zero mean.

The objective of the V-BLAST algorithm is to find $\mathbf{s}$ given $\mathbf{r}$ and $\mathbf{H}$ in a computationally-efficient way. The V-BLAST processing begins with the 1st Tx symbol and proceeds in sequence to the $m$-th symbol. When the optimal ordering procedure is employed, the Tx indexing is changed prior to the processing. The main steps of the algorithm are as follows [1][2]: (1) The interference cancellation step: at the i-th processing step (i.e., when the signal from the i-th transmitter is being detected) the interference from the first (i-1) transmitters can be subtracted based on the estimations of the Tx symbols and the knowledge of the channel matrix $\mathbf{H}$; (2) The inter-stream interference nulling (zero-forcing) step: based on the knowledge of the channel matrix, the interference from yet-to-be-detected symbols (inter-stream interference, ISI) can be nulled out using the orthogonal projection on the sub-spaced orthogonal to that spanned by the yet-to-be-detected symbols; (3) The optimal ordering procedure: the order in which Tx symbols are detected is optimized in such a way that symbols with highest after-processing SNR are detected first. A geometrically-based model of the algorithm, which is used in the present study, has been described in details in [6]. Assuming that the columns of the channel matrix $\mathbf{H} = [\mathbf{h}_1, \mathbf{h}_1, ... \mathbf{h}_m]$ are re-ordered according to the optimal ordering procedure, $\mathbf{H}' = [\mathbf{h}'_1, \mathbf{h}'_1, ... \mathbf{h}'_m]$, the after-processing SNR at i-th step with optimum ZF weights is $\gamma_i = |\mathbf{h}'_{i\perp}|^2 / \sigma_0^2$, where $\mathbf{h}'_{i\perp}$ is the projection of $\mathbf{h}'_i$ on the sub-space orthogonal to $span\{\mathbf{h}'_{i+1}, \mathbf{h}'_{i+2}, ... \mathbf{h}'_m\}$. Fig. 1 illustrates the problem geometry for $m=3$ case. As an example, the optimal ordering at 1st step consists in choosing such $\mathbf{h}'_1 = \mathbf{h}_i$ so that its projection orthogonally to the other two vectors is maximized. The instantaneous block error rate (BLER), i.e. a probability to have at least one error in the detected Tx symbol vector, can be expressed as $P_B = 1 - \prod_{i=1}^{m}(1-P_{ei})$ where $P_{ei} = P_e(\gamma_i)$ is the instantaneous (for given channel realization) error rate at step $i$ conditioned on no errors at steps $1...(i-1)$ [3]-[6]. In the case of unordered V-BLAST, the average (over the channel fading) BLER $\overline{P}_B$ can be simply obtained by using the average step error rate $\overline{P}_{ei}$ in the instantaneous BLER expression (due to the fact that $\gamma_i \sim \chi^2_{2(n-m+i)}$ are independent of each other in i.i.d. Rayleigh fading channel), which is essentially the average error rate of a maximum ratio combiner (MRC) with the appropriate diversity order $(n-m+i)$ [3][5]. For the ordered V-BLAST, however, this does not work any more as $\gamma_i$ are not independent due to the ordering procedure. Furthermore, their statistics is also affected by the optimal ordering. The solution of this problem for the case of $m = 2$ has been given in [6][6]. Here we consider the case of $m \geq 3$.

### III. V-BLAST ALGORITHM ANALYSIS: NxM SYSTEM

In this section, we extend the analysis in [6] to the case of $n \times m$ system, $m > 2$, in an i.i.d. Rayleigh fading channel. This generalization is non-trivial and presents serious mathematical difficulties, which we resolve using various bounds and approximations.

A simple way to obtain a lower bound on the BLER (and also on the outage probability) is to use a genie-assisted system [9], where the genie gives the receiver the symbols of $k$ last transmitters, $k = 1...m-2$,

$$P_B^{(n \times m)} \geq P_B^{(n \times (m-1))} \geq ... \geq P_B^{(n \times 2)} \tag{2}$$

Thus, the BLER of $n \times 2$ V-BLAST $P_B^{(n \times 2)}$ in [6] serves as a lower bound on the BLER of $n \times m$ V-BLAST $P_B^{(n \times m)}$. Note that (2) also applies to the average BLER and holds for arbitrary SNR. Unfortunately, for the average BLER at high SNR, the lower bound $\overline{P}_B^{(n \times 2)}$ is not tight, since its diversity order is $(n-1)$ and the diversity order of $\overline{P}_B^{(n \times m)}$, as it is shown below, is $(n-m+1) < (n-1)$ (for $m > 2$).

To overcome this problem, we use the same geometric model as in [6] (for the $m = 2$ case), whose extension to $m > 2$ results in a generalized form of $1^{st}$ step outage probability $F_1(x)$,

$$F_1(x) = \int_0^{\pi/2} ... \int_0^{\pi/2} f_\varphi(\varphi_1,...\varphi_m) \prod_{i=1}^{m} F_{MRC}^{(n)}\left(\frac{x}{\sin^2 \varphi_i}\right) d\varphi_1...d\varphi_m \tag{3}$$

where $x = \gamma / \gamma_0$ is the SNR normalized to the average one, $\gamma_0 = 1/\sigma_0^2$, $f_\varphi(\varphi_1,...\varphi_m)$ is the joint pdf of $\{\varphi_1,...\varphi_m\}$, $\varphi_i$ being the angle between $\mathbf{h}_i$ and the sub-space spanned all the other column vectors, $f_\varphi(\varphi) = 2(m-1)C_{n-1}^{m-1}\sin^{2(n-m)+1}\varphi \cdot \cos^{2m-3}\varphi$ is the marginal PDF, which can be derived using the same technique as in the $m=2$ case [6], $C_n^m = \frac{n!}{m!(n-m)!}$ is the binomial coefficient, and $F_{MRC}^{(n)}(x) = 1 - e^{-x}\sum_{i=0}^{n-1}x^i/i!$ is the outage probability of n-th order MRC. Note that $f_\varphi(\varphi_1,...\varphi_m)$ is symmetric with respect to $\{\varphi_1,...\varphi_m\}$ (i.e., any two angles can be exchanged without affecting the pdf) due to the problem symmetry. The angles are neither independent nor fully correlated, which makes it very difficult to find the joint pdf required in (3). To this end, we use the Holder inequality in combination with the induction principle to obtain, after some manipulations, the following bounds:

$$F_1(x) \leq \int_0^{\pi/2} f_\varphi(\varphi)\left[F_{MRC}^{(n)}\left(\frac{x}{\sin^2\varphi}\right)\right]^m d\varphi \leq \int_0^{\pi/2} f_\varphi(\varphi) F_{MRC}^{(n)}\left(\frac{x}{\sin^2\varphi}\right) d\varphi \qquad (4)$$

---

[6] The analysis in [6] was based on the ratio of total after-projection signal and noise powers, which is not optimal and implicitly corresponds to the non-coherent (i.e. power-wise) combining after the projection. The results in [6], however, also apply to the optimum zero-forcing weights, with a minor modification only: the step SNR $|\mathbf{h}'_{i\perp}|^2 /(n-m+i)\sigma_0^2$ in [6] should be changed to $|\mathbf{h}'_{i\perp}|^2 / \sigma_0^2$ as in [5].

These inequalities have an intuitive interpretation: the 2nd (rightmost) bound is the 1st step outage probability of unordered V-BLAST (see [3][5] for detailed analysis of this system), and the 1st (internal) bound is the 1st step outage probability of V-BLAST with before-projection ordering (based on $|\mathbf{h}_i|$ rather than $|\mathbf{h}_{i\perp}|$), which is clearly sub-optimal. However, the 1st bound is, as we show later on, quite tight for all SNR. In fact, all the three outage probabilities in (4) exhibit the same diversity order $(n-m+1)$ and differ only by a constant, which is a significant improvement over (2). Note also that for $m=2$ the first bound is sharp, i.e. equals exactly to 1st step outage probability (see eq. 29 in [6]). Additionally, since $\{\varphi_1,...\varphi_m\}$ are exchangeable random variables, which are known to have non-negative correlation [7], the following lower bound holds for arbitrary SNR,

$$\left(F_1^{un}(x)\right)^m = \left(\int_0^{\pi/2} f_\varphi(\varphi) F_{MRC}^{(n)}\left(\frac{x}{\sin^2\varphi}\right) d\varphi\right)^m \leq F_1(x) \qquad (5)$$

where $F_1^{un}(x)$ is the 1st step outage probability of the unordered V-BLAST [5]. Additional lower bounds for $F_1(x)$ can be obtained in the same way as in (2), i.e. $F_1^{(n\times m)} \geq F_1^{(n\times(m-1))} \geq ... \geq F_1^{(n\times 2)}$.

After some lengthy manipulations, the 1st bound in (4) can be presented as:

$$B_1(x) = (-1)^{m-2}(m-1)C_{n-1}^{m-1}\left\{\sum_{l=0}^m \alpha_l\left(J_{3l} + J_{4l}\right)e^{-lx} + \sum_{l=2}^m \alpha_l J_{2l} e^{-lx}\right\} \qquad (6)$$

where $\alpha_l = (-1)^l C_m^l$; $J_{2l}, J_{3l}, J_{4l}$ are polynomials,

$$J_{2l} = (lx)^{n-m+1}\sum_{p=0}^{l(n-1)-n+m-2} a_{pl}(lx)^p, \quad J_{3l} = (-1)^{n+1}(-lx)^{n-m+1}\sum_{p=0}^{m-3} b_p(-lx)^p, \quad J_{4l} = \sum_{p=0}^{n-2} d_p(-lx)^p$$

and $b_p, a_{pl}, d_p$ are numerical coefficients,

$$a_{pl} = \sum_{k=\max[0,m-2-p]}^{m-2} \frac{(-1)^k C_{m-2}^k}{(p+k-m+2)!} \sum_{i=\max[0,p-m+2]}^{l(n-1)-n} c_{i+n,l} l^{-i-n}(k+i)!, \quad b_p = \sum_{k=0}^{p} \frac{(-1)^k C_{m-2}^{k-p+m-2}}{k!} \sum_{i=0}^{m-p-3} \frac{(k+i)!}{(i+p+n-m+2)!}$$

$$d_p = \frac{(-1)^p}{p} \sum_{k=0}^{\min[m-2,n-2-p]} \frac{(-1)^k}{n-k-1} C_{m-2}^k, \quad c_{i,l} = \sum_{\substack{i_1+\ldots+i_l=i \\ 0 \le i_1,\ldots,i_l \le n-1}} \frac{1}{i_1!\ldots i_l!}$$

Thus, the bound in (6) is a combination of exponents and polynomials of finite order and hence allows efficient numerical evaluation, which is important, for example, in an iterative optimization process. While the expression for $a_{pl}, b_p, d_p$ may appear complicated, they can be evaluated in advance (i.e., a table of coefficients is built for a given order of the system) and do not need to be changed during simulations. We also note that for $m=2$, (6) reduces to the exact 1st step outage probability (see [[6], eq. 30]), i.e. the bound is sharp in this case.

At high SNR, $B_1(x)$ can be approximated as

$$B_1(x) \approx \frac{1}{(n-m+1)!} \left(\frac{x}{2}\right)^{n-m+1} \quad (7)$$

which indicates a 3 dB SNR gain in the sub-optimally ordered V-BLAST compared to the unordered one. To get some insight and to evaluate the bound accuracy in other cases, we further consider 3x3 and 4x4 systems.

A. *Outage of 3x3 and 4x4 V-BLAST*

The 1st step outage of 3x3 V-BLAST is bounded as

$$F_1(x) \le B_1(x) = 1 - 3e^{-x} + e^{-2x}\left(3 + \frac{15}{8}x + \frac{3}{8}x^2\right) - e^{-3x}\left(1 + \frac{110}{81}x + \frac{7}{9}x^2 + \frac{2}{9}x^3 + \frac{1}{36}x^4\right) \quad (8)$$

The asymptotic behavior of the bound is $B_1(x) \approx x/2, \; x \to 0$, which is the same as the asymptotic outage probability of the 2x2 system [6]. 1st order diversity and 3 dB gain due to optimal ordering are apparent (this 3 dB gain transforms asymptotically into 3 dB SNR gain in terms of the average BLER).

The 2nd step outage of 3x3 V-BLAST cannot be easily evaluated since the ordering procedure at the 1st step affects the channel statistics at the 2nd step. We evaluate the conditional outage probability at the 2nd step (i.e., conditioned on no detection error at the 1st step – this is what we need to evaluate the BLER). As an approximation, we assume that the channel statistics at the 2nd step are not affected by the optimal ordering at the 1st step (i.e., the channel coefficients are still i.i.d. complex Gaussian). Under this assumption, the 2$^{nd}$ step outage probability is the same as that of a 3x2 system at the 1$^{st}$ step (since the first bit stream has been detected and eliminated at the 1$^{st}$ step), whose outage probability is [7, eq. 21], $F_2(x) = 1 - 2e^{-x}(1+x) + e^{-2x}(1+2x+9x^2/8+x^3/4)$. Its asymptotic behavior is $F_2(x) \approx x^2/8$, $x \to 0$, which clearly indicates a second-order diversity.

The 3$^{rd}$ step conditional outage probability can be evaluated in a similar way. Assuming no change in the channel statistics due to the ordering in the first two steps, it is the same as that of a 3x2 system at the second step, $F_3(x) = F_{MRC}^{(3)}(x)\left[2 - F_{MRC}^{(3)}(x)\right]$. Its asymptotic behavior is $F_3(x) \approx 2F_{MRC}^{(3)}(x) \approx x^3/3$, which indicates the 3$^{rd}$ order diversity.

Extensive Monte-Carlo simulations have been carried out to evaluate the accuracy of the bound and approximations involved. Some of the representative results are shown in Fig. 2-4. Clearly, the 1$^{st}$ step bound is quite accurate (given its simple nature) and it underestimates the performance by 2 dB. The actual asymptotic behavior of the outage probability is $F_1(x) \approx x/3$, $x \to 0$. The 2$^{nd}$ step performance is overestimated by 3 dB. However, as Fig. 2 demonstrates, it is predicted extremely well by the 2$^{nd}$ order MRC outage curve. We attribute this to the joint effect of two opposite factors: 1) performance loss at the 2$^{nd}$ step due to optimal ordering at the 1$^{st}$ one (the same as for nx2 system), and 2) performance improvement due to the 2$^{nd}$ step optimal ordering. Apparently, this two effects compensate each other and the resulting outage is the same as that of 2$^{nd}$ order MRC. The 3$^{rd}$ step performance is estimated quite accurately by the approximate expression $F_3(x) \approx 2F_{MRC}^{(3)}(x) \approx x^3/3$ (within 1 dB). MRC outage curve would provide worse approximation in that case.

The validity of the approximations above is not limited to a 3x3 system. As an example, we use the same approximations to analyze 4x4 system. Fig. 3 shows the outage probability at first 3 steps. The 1$^{st}$ step bound is obtained using (6), and the asymptotic behavior of the outage probability, based on Monte-Carlo simulations, is $F_1(x) \approx x/4$, $x \to 0$ (see (9) for $m=4$). The 2$^{nd}$ step outage has been analytically estimated using the 1$^{st}$ step outage of a 4x3 system, which is within 1.5 dB of the actual performance. Note that it is not the same as MRC anymore. However, the 3$^{rd}$ step performance is virtually the same as that of 3$^{rd}$ order MRC. The analytic estimation of the performance (using 1$^{st}$ step outage of a 4x2 system) overestimates it by approximately 3 dB. We attribute this to the effect of the optimal ordering at the 1$^{st}$ and 2$^{nd}$ steps.

B. *Outage of n×m V-BLAST*

Extensive numerical experiments (Monte-Carlo simulations) indicate that the bound in (7) captures accurately the diversity order ($=n-m+1$), but underestimates the SNR gain of the optimal ordering: instead of 2, the SNR gain observed in numerical experiments for systems of various sizes is $m$ and the outage probability is accurately approximated, at high SNR, by

$$F_1(x) \approx \frac{1}{(n-m+1)!}\left(\frac{x}{m}\right)^{n-m+1}, \quad x \to 0 \tag{9}$$

Some of the representative results shown in Fig. 4 demonstrate that (9) is indeed accurate for systems of various sizes. When compared to the $(n-m+1)$-order MRC outage probability, $F_{MRC}^{(n-m+1)}(x) \approx x^{n-m+1}/(n-m+1)!$, which is also the 1$^{st}$ step outage of the un-ordered V-BLAST [3]-[5], it is clear that the effect of the optimal ordering is an $m$-fold SNR gain at the 1$^{st}$ step. This was rigorously proved for $n \times 2$ system in [6]. Fig. 5 shows additional representative results in terms of the average BLER with BPSK modulation evaluated via (9) (see also (11) and (12)). Clearly, this approximation captures accurately not only the diversity order, but also the SNR gain of the optimal ordering, at high SNR. Another related approximation, also supported by extensive numerical evidence, gives the joint distribution $F(x_1, x_2, ..., x_m)$ of $\{|\mathbf{h}_{1\perp}|^2, |\mathbf{h}_{2\perp}|^2, ..., |\mathbf{h}_{m\perp}|^2\}$ [16],

$$F(x_1, x_2, ..., x_m) \approx \frac{1}{(n-m+1)!}\left(\sum_{i=1}^{m} x_i^{-1}\right)^{n-m+1}, \quad x_i \to 0 \tag{10}$$

where $\mathbf{h}_{i\perp}$ denotes here the component of $\mathbf{h}_i$ orthogonal to $span\{\mathbf{h}_1...\mathbf{h}_{i-1}, \mathbf{h}_{i+1}...\mathbf{h}_m\}$. Note that (9) follows from (10), since $F_1(x) = F(x, x, ..., x)$.

Comparing (9) to (7), we conclude that the sub-optimal ordering (based on $|\mathbf{h}_i|$) incurs $10\log_{10}(m/2)$ dB SNR penalty compared to the optimal ordering (based on $|\mathbf{h}_{i\perp}|$). This SNR gap is zero for $m=2$ and it widens as $m$ increases. However, the sub-optimal ordering does provide a 3 dB SNR gain compared to the unordered V-BLAST. Given a lower computational complexity of the sub-optimally ordered V-BLAST[7], it may be an attractive solution for low-complexity or energy-limited systems.

IV. AVERAGE BLER AND TBER

Using the analytical approximations for the outage probabilities given above, the average BLER can be evaluated in a straightforward way. Since, in large SNR regime, $\overline{P_{e,1}} \gg \overline{P_{e,2}} \gg ... \gg \overline{P_{e,m}}$ (due to increasing diversity order), the average BLER is dominated by the 1st step average error rate. Using (9), this can be presented as

$$\overline{P_B} \approx \overline{P_{e,1}} \approx \overline{P}_{MRC}^{(n-m+1)}(m\gamma_0) \tag{11}$$

where $\overline{P}_{MRC}^{(k)}(\gamma_0)$ is the k-th order MRC average BER for a given modulation format. Specifically, for non-coherent orthogonal BFSK and coherent BPSK respectively one obtains,

$$\overline{P}_{B,BFSK} \approx \frac{1}{2}\left(\frac{2}{m\gamma_0}\right)^{n-m+1} \quad , \quad \overline{P}_{B,BPSK} \approx \frac{C_{2(n-m)+1}^{n-m+1}}{(4m\gamma_0)^{n-m+1}} \tag{12}$$

---

[7] for example, one projection is required at 1st step instead of $m$ for the optimal ordering; given that the projections carry most of the V-BLAST computational load, the reduction in complexity can be significant.

To evaluate the validity and accuracy of the approximations in (11), (12) extensive Monte-Carlo simulations have been carried out. Some of the representative results are shown in Fig. 5, 6[8]. While the overall accuracy of these approximations is good, we note that the large-SNR approximation in (12) is less accurate at smaller SNR (especially when applied to $\overline{P_{e,1}}$) compared to the general expression in (11) (which is not a surprise as (12) follows from (11), which is itself a large-SNR approximation). We have noticed that a better approximation for the average BLER is

$$\overline{P_B} \approx \overline{P}_{MRC}^{(n-m+1)}(m\gamma_0) + \overline{P}_{MRC}^{(n-m+2)}(\gamma_0) \tag{13}$$

where the 2nd term approximates the contribution of the 2nd and higher steps, whose contribution to the average BLER is visible at lower SNR. For large average SNR, $\gamma_0 \geq 10dB$, all these approximations are almost identical, which also confirms our earlier conclusion that the 1st step has dominant effect on the average BLER. No diversity for 3x3 and 2nd order diversity for 4x3 system is also obvious (both, in terms of BLER and 1st step BER). For $m \times m$ system, (11) simplifies to

$$\overline{P_B} \approx \frac{a}{m\gamma_0} \tag{14}$$

where the constant *a* is independent of *m*, from which one may conclude that

$$\lim_{m \to \infty} \overline{P_B} = 0 \tag{15}$$

While we lack a rigorous proof of this result (recall that it is based on the conjecture (9)), there is an extensive numerical evidence to support it. As Fig. 6 demonstrates for BPSK modulation, the average BLER decreases with *m* in complete agreement with (14). Note also that the approximation in (14) becomes quite accurate for high SNR, $\gamma_0 > 10dB$. As a side remark, we also note that this approximation over-estimates the average BLER. From (15), one may conclude that the V-BLAST is a practical example of a system with the space-time autocoding effect discovered in [8] using information-theoretic arguments. Together with the capacity-achieving property of the V-BLAST [1], this demonstrates the optimality of its processing strategy.

---

[8] $10^6$ channel realizations with 100 noise/symbol realization per channel realization have been used for the MC simulations. The rationale behind this choice can be found in [15].

Another performance measure of the BLAST used in the current literature is the total error rate (TBER). It is defined as the error rate at the output stream to which all the individual sub-streams are merged after the detection. Thus, contrary to the BLER, it takes into account the actual number of errors at the transmitted symbol vector and not only the fact of their presence. The two measures are related via the following inequalities [5]: $\frac{1}{m}\overline{P}_B \leq \overline{P}_T \leq \overline{P}_B$, where $\overline{P}_T$ is the average TBER. The lower bound is tight at high SNR, so it can be used as an accurate approximation of the average TBER [5], $\overline{P}_T \approx \frac{1}{m}\overline{P}_B$, i.e. single errors within the Tx symbol vector dominate the performance at high SNR. Using this approximation and (11), the average TBER of the ordered V-BLAST in the i.i.d. Rayleigh fading channel can be approximated as

$$\overline{P}_T \approx \frac{1}{m} \overline{P}_{MRC}^{(n-m+1)}(m\gamma_0) \tag{16}$$

Thus, increasing $m$ for fixed $n$ results in two opposing effects on the average TBER: (i) decreasing diversity order, and (ii) increasing SNR gain due to the ordering and dominance of single-error events. Unfortunately, at high SNR, the 1st (negative) effect is dominant. However, when $m$ and $n$ increase simultaneously, so that $(n-m)$ is fixed, the 1st effect disappears, and the second (positive) effect dominates the performance. This observation corroborates the common wisdom that an increase in the number of Tx antennas should be followed by an appropriate increase in the number of Rx antennas, if the error rate performance must not degrade. It also demonstrates that, for $m = n$, an increase in $m$ is beneficial in terms of the average TBER. For example, when $n = m$, (16) reduces to

$$\overline{P}_T \approx \frac{a}{m^2 \gamma_0} \tag{17}$$

Comparing (17) to (14), one concludes that the auto-coding effect is even more pronounced in terms of the average TBER. In a similar way, one may express the average TBER for BFSK and BPSK using (16) and (12).

V. D-BLAST AND COMPARISON TO LINEAR INTERFACES

The error rate performance, either instantaneous or average, of the uncoded D-BLAST (ordered or not) is the same as that of the V-BLAST [5], since antenna cycling in D-BLAST is equivalent to symbol cycling at the baseband model, so if the same constellation is used for each transmitter, the total error rate is not affected. In the case of the BLER, the error probability is not affected by cycling either. Thus, the results above can also be applied to the D-BLAST algorithm. In this case the step index $i$ in Section II is associated with the antenna, not the transmitter. The identical performance, however, does not hold for horizontally-coded BLAST, in which case there is a significant difference, with D-BLAST outperforming V-BLAST [9].

The BLAST algorithm performs essentially successive interference constellation (SIC). Another option would be to use purely linear processing, i.e. either ZF or MMSE filtering applied to $\mathbf{r}$ to form decision variables as follows [10, sec. 10.3]: $\mathbf{r}'_{ZF} = (\mathbf{H}^+\mathbf{H})^{-1}\mathbf{H}^+\mathbf{r}$, $\mathbf{r}'_{MMSE} = (\mathbf{H}^+\mathbf{H} + \sigma_0^2\mathbf{I})^{-1}\mathbf{H}^+\mathbf{r}$, and the demodulation is done component-wise, $\hat{x}_i = D^{-1}\{r'_i\}$, where $D^{-1}$ denotes a demodulator for a constellation in use. Clearly, each stream will experience in this case the same error rate (due to the same distribution of the after-processing SNR) equal to the 1$^{st}$ step error rate of unordered V-BLAST with ZF and MMSE processing respectively. Thus, the average BLER of the linear interface can be expressed as,

$$\overline{P}_B^L = 1 - \prod_{i=1}^{m}(1 - \overline{P}_{e1}^u) \approx m\overline{P}_{e1}^u \tag{18}$$

where $\overline{P}_{e1}^u$ is the 1$^{st}$ step error average rate of the unordered V-BLAST (for ZF processing, corresponding expressions were given in [5]; for MMSE processing, it seems to be an open problem), and the approximate equality holds at large SNR, when $\overline{P}_{e1}^u \ll 1$. Comparing this to the average BLER of unordered V-BLAST at high SNR, $\overline{P}_B^u \approx \overline{P}_{e1}^u$, one concludes that the linear interface suffers $m$-fold increase in the average BLER compared to the SIC, which is equivalent to $m^{\frac{1}{n-m+1}}$-fold SNR penalty. For the ordered ZF V-BLAST, the average BLER is given by (11), and $\overline{P}_{e1}^u \approx \overline{P}_{MRC}^{(n-m+1)}(\gamma_0)$ [5], so this difference is even more pronounced since

$$\overline{P}_B^L \approx m\overline{P}_{MRC}^{(n-m+1)}(\gamma_0) > \overline{P}_B^u \approx \overline{P}_{MRC}^{(n-m+1)}(\gamma_0) > \overline{P}_B \approx \overline{P}_{MRC}^{(n-m+1)}(m\gamma_0) \tag{19}$$

For the $m \times m$ system at high SNR, $\overline{P}_{MRC}^{(1)}(m\gamma_0) \approx mc_1/\gamma_0$, where $c_1$ is a modulation-depend constant, so that (19) reduces to

$$\overline{P}_B^L \approx mc_1/\gamma_0 > \overline{P}_B^u \approx c_1/\gamma_0 > \overline{P}_B \approx c_1/(m\gamma_0) \tag{20}$$

i.e. the linear interface suffers $m^2$-fold SNR penalty compared to the ordered V-BLAST. For the $n \times m$ system, this becomes $m^{\frac{n-m+2}{n-m+1}}$-fold SNR penalty. This simplified analysis agrees reasonably well with the numerical results in [10, sec. 10.3].

## VI. CONCLUSIONS

Geometrically-based analytical performance evaluation of the ordered ZF V-BLAST for more than 2 Tx antennas is presented. The use of the Holder inequality allows one to derive lower and upper bounds on the $1^{st}$ step outage probability at arbitrary SNR, and also compact closed-form approximations for higher-steps outage probabilities at high SNR. Since the average BLER and TBER are dominated by the $1^{st}$ step BER at that mode, approximate closed-form expressions are obtained in a straightforward way. Based on the upper bound to the outage probability and on extensive numerical evidence, we conclude that the effect of optimal ordering at $1^{st}$ step is $m$-fold SNR gain. Thus, the average BLER $\rightarrow 0$ (and also the TBER) as $m \rightarrow \infty$, which is a manifestation of the space-time autocoding effect [8] in a practical system, which also confirms the optimality of the V-BLAST processing strategy. Contrary to the optimally-ordered V-BLAST, the sub-optimally ordered one and un-ordered one possess the autocoding effect in terms of the TBER but not the BLER. Uncoded D-BLAST error rate performance is the same as that of the V-BLAST. The linear receiver interface (without successive interference cancellation) suffers SNR penalty compared to the BLAST, which is quantified in this paper.

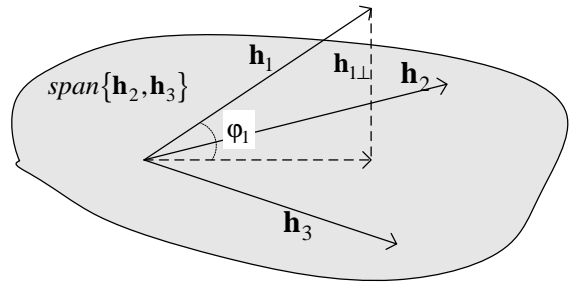

Fig. 1. Illustration of the problem geometry for $m = 3$.

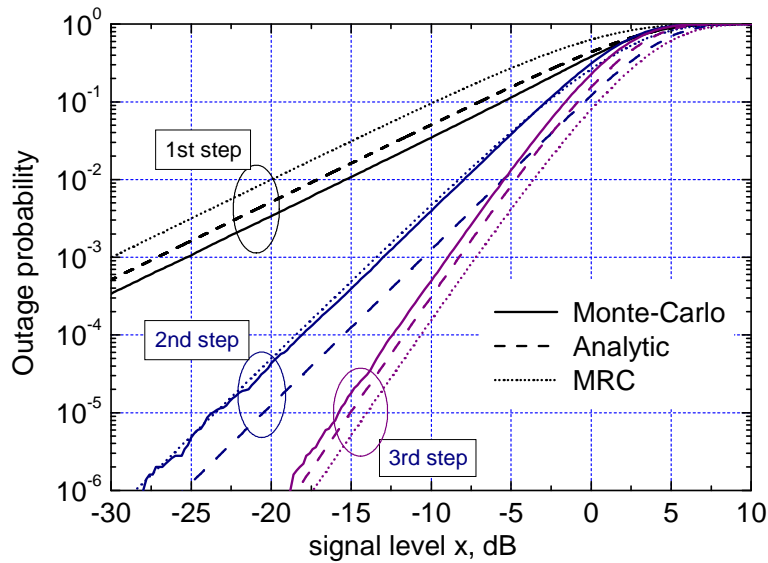

Fig. 2. Outage probabilities of 3x3 optimally-ordered V-BLAST. $5*10^6$ trials have been used for Monte-Carlo simulations.

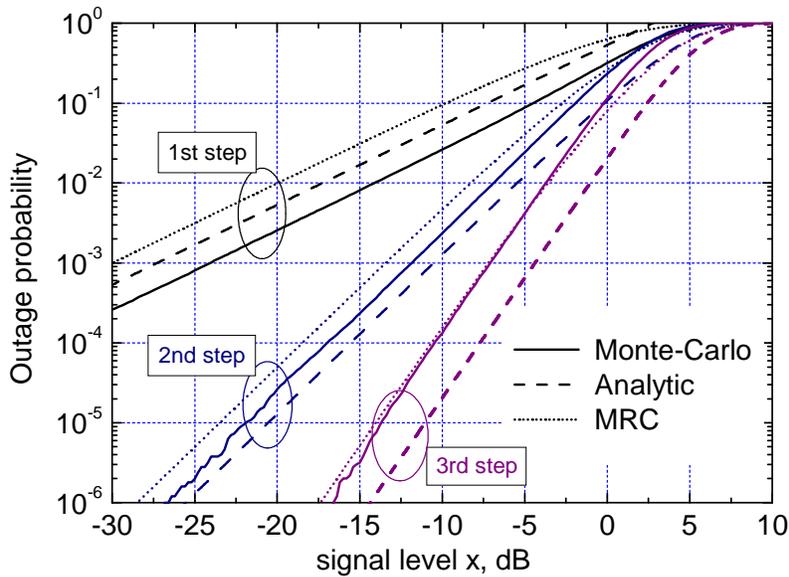

Fig. 3. Outage probabilities of 4x4 optimally-ordered V-BLAST. $5*10^6$ trials have been used for Monte-Carlo simulations.

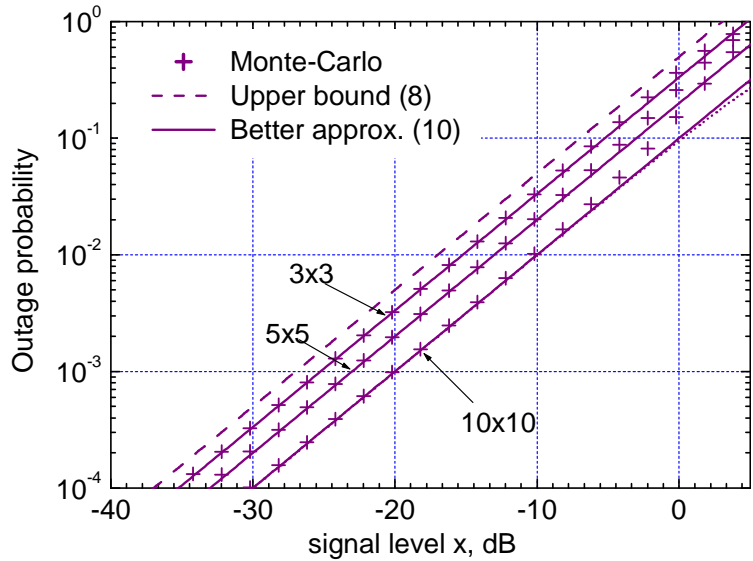

Fig. 4. $1^{st}$ step outage probabilities of optimally-ordered V-BLAST of various sizes evaluated via (7), (9) and via Monte-Carlo simulations. While the bound in (7) captures correctly the diversity order, it gets less and less tight as $m$ increases and thus unable to capture accurately the SNR gain of ordering. The approximation in (9) captures accurately both the diversity order and the SNR gain of the ordering, at high SNR. It also allows estimating accurately the BLER (see Fig. 5 and 6).

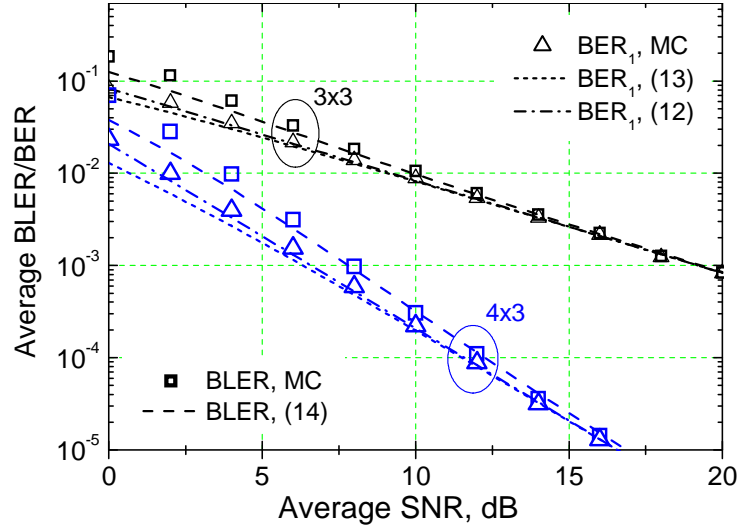

Fig. 5. Average BER/BLER of 3x3 and 4x3 optimally-ordered V-BLAST with BPSK modulation: approximations and MC simulations. The approximations become accurate at about $SNR \geq 5dB$.

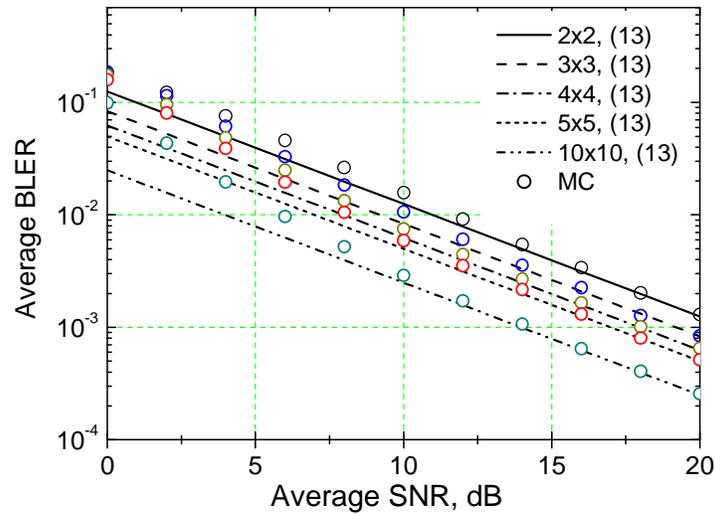

Fig. 6. Average BLER of mxm optimally-ordered V-BLAST with BPSK modulation using the high-SNR approximations in (11) and (12), and MC simulations for $m = 2, 3, 4, 5$ and $10$. The approximation becomes very accurate at about $SNR \geq 10dB$.